\documentclass[final,onefignum,onetabnum]{siamart190516}

\usepackage{braket,amsfonts}  
\usepackage{amssymb}  
\usepackage{mathtools}
\usepackage{multicol}
\usepackage{multirow}
\usepackage{algorithmic}
\usepackage{xcolor}
\usepackage[normalem]{ulem}

\DeclareMathOperator*{\argmin}{arg\,min}

\title{Utilizing variational autoencoders in the Bayesian inverse problem of photoacoustic tomography \thanks{\funding{This project has received funding from the European Research Council (ERC) under the European Union’s Horizon 2020 research and innovation programme (grant agreement No 101001417- QUANTOM). This work was supported by the Academy of Finland (projects 314411, 336799 Centre of Excellence in Inverse Modeling and Imaging,  and 320166 the Flagship Program Photonics Research and Innovation ), and the Jane and Aatos Erkko Foundation.}}}

\author{Teemu Sahlström\thanks{University of Eastern Finland, Department of Applied Physics}
	\and Tanja Tarvainen\thanks{University of Eastern Finland, Department of Applied Physics; University College London, Department of Computer Science}}

\begin{document}
	\maketitle
	
	\begin{abstract}
		There has been an increasing interest in utilizing machine learning methods in inverse problems and imaging. Most of the work has, however, concentrated on image reconstruction problems, and the number of studies regarding the full solution of the inverse problem is limited. In this work, we study a machine learning based approach for the Bayesian inverse problem of photoacoustic tomography. We develop an approach for estimating the posterior distribution in photoacoustic tomography using an approach based on the variational autoencoder. The approach is evaluated with numerical simulations and compared to the solution of the inverse problem using a Bayesian approach. 
	\end{abstract}

	
	
	\section{Introduction}
	\label{sec:Introduction}
	
	Photoacoustic tomography (PAT) is a hybrid biomedical imaging modality based on the photoacoustic effect \cite{Beard2011, Wang2016, Poudel2019}. In PAT, the imaged target is illuminated with a short pulse of light. Absorption of light creates localized areas of thermal expansion, resulting in localized pressure increases within the imaged target. This pressure distribution, called the initial pressure, relaxes as broadband ultrasound waves that are measured on the boundary of the imaged target. In the inverse problem of PAT, the initial pressure distribution is estimated from a set of measured ultrasound data. 
	
	Various methods for reconstruction of photoacoustic images have been used \cite{Poudel2019}. These include, for example, analytical methods \cite{Haltmeier2014,Haltmeier2007,Haltmeier2005,Agranovsky2018,Kunyansky2007,Wang2005,Finch2004,KuchmentBook}, the time-reversal \cite{Xu2004, Treeby2010, Burgholzer2007, Hristova2008}, regularized least-squares approaches \cite{Arridge2016,Deanben2012,Wang2012,Wang2013}, and the Bayesian approach \cite{Tick2016,Tick2018,Sahlstrom2020}. Most of the image reconstruction methods, such as the regularized least-squares, provide point estimates of the estimated initial pressure distribution. Using suitable regularization, these methods can be used to alleviate various artifacts and improve overall image quality. These approaches do not, however, offer information regarding the reliability of the estimated image. In the Bayesian framework, the inverse problem of PAT is approached in a framework of statistical inference. The solution of the inverse problem is the posterior distribution, that is a conditional probability distribution of the unknown parameters, given measurements, a forward model and a prior model for the unknowns. The Bayesian approach facilitates representing and taking into account uncertainties in parameters, models and measurement geometries \cite{KaipioBook,Arridge2016,Sahlstrom2020,Tick2019}.
	
	In recent decades, utilization of machine learning and deep learning techniques in various imaging modalities has increased tremendously. In PAT, machine learning has been applied to a wide range of problems, see e.g. the recent reviews \cite{MLReviewHauptmann2020,MLReviewGrohl2021,MLReviewYang2020} and the references therein. The first machine learning methods utilized in PAT can be divided as post- and pre-processing methods. In the post-processing approaches, a photoacoustic image is first reconstructed using a conventional reconstruction method. Then, a neural network is used to correct, for example, limited view artifacts or noise in the image \cite{Antholzer2018,Guan2020,Davoudi2019}. In the pre-processing approaches, a photoacoustic dataset is first processed using a neural network to improve the signal quality and an image is subsequently reconstructed using a conventional approach \cite{Davoudi2019,Allman2018}. In addition to the pre- or post-processing, deep learning has been applied in the reconstruction process to fully or partially replace the conventional reconstruction procedure. The simplest of these approaches is the so-called end-to-end framework, where a neural network is trained to perform the reconstruction, that is, to estimate the initial pressure distribution based on a set of photoacoustic data \cite{Waibeletal2018,Lan2019}. These networks are, however, often trained by minimizing the difference between the network output and the true image contained in a training dataset. Therefore, the results might be inconsistent with measurement data and may lack the ability to generalize to targets outside the training dataset. 
	
	To overcome the issues with data consistency, approaches utilizing forward and adjoint operators during the training process have been proposed \cite{Lunz2021}. In these learned iterative model based approaches, the inverse problem of PAT is solved iteratively with a learned updating operator. The approach has been been utilized in an iterative proximal gradient based framework where an updating operator was learned based on a previous iterate and the gradient of a data consistency term \cite{Shan2019,Hauptmann2018}. Furthermore, in a learned primal dual method, two learned update operators are used to update the iterations in primal (image) and dual (data) spaces \cite{Boink2019}. In addition to iterative update operators, forward operators together with deep learning in PAT have been used in conjunction with least-squares type approaches with a learned regularization \cite{Antholzer2019}. In other optical and ultrasonic imaging modalities, forward operators and deep learning have been utilized in model correction similarly as in Bayesian approximation error modeling \cite{Mozumder2022, Koponen2021}.
	
	As with the conventional reconstruction approaches, most of the machine learning based approaches provide images of the underlying initial pressure distribution but do not offer insight in quantifying the reliability of the solution. Previously, uncertainties in PAT in the context of neural networks have been studied using a dropout Monte Carlo (MC) \cite{Godefroy2021}. In this approach, dropout layers are used while computing several outputs of a neural network to provide a set of images. From this set of images, the mean and standard deviation can be computed to evaluate the uncertainty of the network. That is, MC dropout aims to quantify the uncertainty of the neural network model, by introducing variability via random deactivation of network layers. Furthermore, uncertainties in quantitative PAT using machine learning have been studied by considering aleatoric and epistemic sources of uncertainties \cite{Grohl2018}. In that work, an external observing neural network was used to estimate the uncertainty of the network which was subsequently combined with an estimate of the random uncertainty to quantify the uncertainties in the reconstruction.
	
	In this work, we propose a machine learning based approach to the Bayesian inverse problem of PAT. The approach is based on the variational autoencoder (VAE) \cite{Kingma2014} and the recently proposed extension to the VAE called the uncertainty quantification VAE (UQ-VAE) \cite{Goh2021}. Conventional implementations of the VAE consist of an encoder and decoder that output the parameters to an approximate posterior distribution and data likelihood distribution, respectively. In the proposed framework, however, the decoder is replaced by an explicitly computed data likelihood term facilitating the use of the true forward operator during the training process. Furthermore, compared to the unsupervised framework of the VAE, that is trained solely on a dataset of measurement data, the UQ-VAE provides the possibility of utilizing the underlying initial pressure images in the training process. Once the network parameters have been trained, the UQ-VAE enables estimating the posterior distribution from a (single) photoacoustic dataset. Thus, it provides photoacoustic images together with estimates of their reliability taking into account uncertainties arising from the measurement noise, forward model and prior distribution.
	
	The remainder of this article is structured as follows. The forward problem and the Bayesian framework to the inverse problem of PAT are described in Section \ref{sec:Photoacoustic tomography}. Framework for solving the Bayesian inverse problem using the VAE and UQ-VAE are presented in Section \ref{sec:Uncertainty quantification using neural networks}. Simulations are described in Section \ref{sec:Simulations}, and their results are presented in Section \ref{sec:Results}. Finally, the results are discussed and conclusions are given in Sections \ref{sec:Discussion} and \ref{sec:Conclusions}.

	\section{Photoacostic tomography}
	\label{sec:Photoacoustic tomography}
	
	\subsection{Forward model}
	
	Ultrasound propagation generated by an initial pressure $p_0(r)$ in an acoustically homogeneous and non-attenuating medium can be described as an initial value problem 
	\begin{align}
		\begin{dcases}
			\nabla^2 p(r,t) - \frac{1}{c^2} \frac{\partial^2p(r,t)}{\partial t^2} = 0 \\
			p(r,t = 0) = p_0(r) \\
			\frac{\partial}{\partial t} p(r,t=0) = 0,
		\end{dcases}
		\label{eq:Wave_equation}
	\end{align}
	where $\nabla^2$ is the Laplace operator,  $p(r,t)$ is the pressure at a point $r$ and time instance $t$, and $c$ is the speed of sound. In practice, the measured pressure denoted by $p_t$, is measured on a finite number of points or surfaces around the imaged target. In this work, the solution of \eqref{eq:Wave_equation} is numerically approximated using a pseudospectral $k$-space method implemented within the k-Wave MATLAB toolbox \cite{Treeby2010kwave}.
	
	\subsection{Inverse problem}
	\label{sec:PAT_InverseProblem}
	
	Let us consider a measurement geometry consisting of $N$ ultrasound sensors, a spatial discretization consisting of $H$ pixels and a temporal discretization of $M$ time points. The discrete observation model for PAT can be written as
	\begin{equation}
		p_t = Kp_0 + e,
		\label{eq:forwMod}
	\end{equation}
	where $p_t \in \mathbb{R}^{NM\times 1}$ is a vector of measured photoacoustic data, $K \in \mathbb{R}^{MN \times H}$ is a discrete forward operator, $p_0 \in \mathbb{R}^{H \times 1}$ is a vector of the initial pressure, and $e \in \mathbb{R}^{MN \times 1}$ is additive measurement noise. 
	
	In this work, the inverse problem of PAT is approached in a Bayesian framework \cite{KaipioBook,Tick2016}. The solution of the inverse problem is the posterior distribution given by the Bayes' formula \cite{KaipioBook}
	\begin{equation}
		\pi(p_0 \vert p_t) \propto \pi(p_t \vert p_0) \pi(p_0),
	\end{equation}
	where $\pi(p_t \vert p_0)$ is the data likelihood and $\pi(p_0)$ is the prior distribution. Let us model the initial pressure $p_0$ and the measurement error $e$ as mutually independent and Gaussian distributed $p_0 \sim \mathcal{N}(\eta_{p_0}, \Gamma_{p_0})$, $e \sim \mathcal{N}(\eta_{e}, \Gamma_{e})$, where $\eta$ and $\Gamma$ are the mean and covariance of the respective distributions. Then, the data likelihood can be written as a Gaussian distribution \cite{KaipioBook}
	\begin{equation}
		\pi(p_t \vert p_0) \propto \exp \left\{ -\frac{1}{2} \left\Vert L_e(p_t - Kp_0 - \eta_e) \right\Vert^2_2 \right\},
		\label{eq:ExpLikelihood}
	\end{equation}
	where $L_e$ is the inverse of the square root of the noise covariance matrix such as the Cholesky decomposition  $\Gamma_e^{-1} = L_e^{\mathrm{T}}L_e$. Now, in the case of a linear forward model and Gaussian distributed noise and prior, the posterior distribution is a Gaussian distribution \cite{KaipioBook}
	\begin{equation}
		\pi(p_0 \vert p_t) \propto \exp \left\{ -\frac{1}{2} \left\Vert L_e(p_t - Kp_0 - \eta_e) \right\Vert^2_2 -\frac{1}{2} \left\Vert L_{p_0}(p_0 - \eta_{p_0}) \right\Vert^2_2  \right\},
		\label{eq:Posterior_exp}
	\end{equation}
	where $L_{p_0}$ is the Cholesky decomposition of the prior covariance  $\Gamma_{p_0}^{-1} = L_{p_0}^{\mathrm{T}}L_{p_0}$. The mean $\eta_{p_0 \vert p_t}$ and covariance $\Gamma_{p_0 \vert p_t}$ of the posterior distribution $\pi(p_0 \vert p_t)$ are \cite{KaipioBook,Tick2016}
	\begin{align}
		\eta_{p_0 \vert p_t} &= (K^\mathrm{T}\Gamma_e^{-1}K + \Gamma_{p_0}^{-1})^{-1}(K^\mathrm{T}\Gamma_e^{-1}(p_t - \eta_e) + \Gamma_{p_0}^{-1}\eta_{p_0}) \label{eq:Posterior_mean}\\
		\Gamma_{p_0 \vert p_t} &= (K^\mathrm{T}\Gamma_e^{-1}K + \Gamma_{p_0}^{-1})^{-1}.
		\label{eq:Posterior_cov}
	\end{align}
	In this work, the posterior distribution \eqref{eq:Posterior_exp}-\eqref{eq:Posterior_cov} is referred to as the solution of the inverse problem in the Bayesian framework.

	\section{Approximating posterior distribution using neural networks}
	\label{sec:Uncertainty quantification using neural networks}
	
	\subsection{Variational autoencoder}
	\label{subsec:Auto-Encoding Variational Bayes}
	
	Generative modeling can be described as a task of learning a probability distribution of a random variable based on a set of observations. This data distribution can then be utilized in generating previously unseen data samples. Some of the most utilized generative modeling frameworks include variational autoencoders (VAE) \cite{Kingma2014}, generative adversarial networks (GAN) \cite{Goodfellow2014} and normalizing flows (NF) \cite{Brubaker2020}. 
	
	Even though the generative models such as the VAE are most often used for data generation, they can also be utilized in estimation of posterior distributions. The VAE can be regarded as a variational Bayesian approach, where one aims to seek an approximation to a true intractable posterior distribution from a family of tractable distributions, e.g. Gaussian distributions. This can be accomplished by minimizing the Kullback-Leibler divergence (KLD) between the true posterior distribution $\hat{\pi}(p_0 \vert p_t)$ and an approximation of the true posterior distribution $\pi_\phi(p_0 \vert p_t)$ parameterized by $\phi$. The KLD can be minimized by maximizing the so-called evidence lower bound (ELBO) \cite{Kingma2014}
	\begin{equation}
		\mathcal{L}(\phi) = -\mathrm{KL}(\pi_\phi(p_0 \vert p_t) \Vert \pi(p_0) ) + \mathbb{E}_{p_0 \sim \pi_\phi(p_0 \vert p_t)} \left[ \log \pi(p_t \vert p_0) \right],
		\label{eq:ELBO}
	\end{equation}
	where $\pi(p_0)$ is the prior, $\pi(p_t \vert p_0)$ is the data likelihood as defined in Section \ref{sec:PAT_InverseProblem}, and $\mathbb{E}$ denotes the expected value. 
	
	Let us now consider the linear observation model with additive noise \eqref{eq:forwMod}. Furthermore, let us model the initial pressure $p_0$ and noise $e$ as mutually independent and Gaussian distributed leading to Gaussian approximate posterior distribution $\pi_\phi(p_0 \vert p_t)$. In accordance with this, we define a dataset $P_t^{\textsc{VAE}} = \{p_t^{(i)}, \sigma_e^{(i)}\}_{i=1\dots,I}$ consisting of pairs of photoacoustic measurement data $p_t^{(i)}$ and the standard deviations of the noise $\sigma_e^{(i)}$. In $P_t^{\textsc{VAE}}$, the measurement data $p_t^{(i)} \in \mathbb{R}^{NM \times 1}$ constitutes a noisy vectorized photoacoustic dataset with the noise characterized by the standard deviation $\sigma_e^{(i)}$ of the additive Gaussian noise.
	
	Let us further define a neural network $\Lambda$, that takes a sample of the data $p_t^{(i)}$ as an input and outputs the statistics, i.e. the mean $\eta_{p_0 \vert p_t}^{(i)}$ and covariance $\Gamma_{p_0 \vert p_t}^{(i)}$, of the approximate posterior distribution $\pi_\phi(p_0 \vert p_t)$. Then, using the so-called reparameterization trick \cite{Kingma2014}, the ELBO \eqref{eq:ELBO} for a data sample $p_t^{(i)}$ can be written as

	\begin{equation}
		\mathcal{L}(\phi; p_t^{(i)}) = 
		-\mathrm{KL}(\pi_\phi(p_0 \vert p_t^{(i)}) \Vert \pi(p_0) ) +  \log \pi(p_t^{(i)} \vert \tilde{p}_0^{(i)}),
		\label{eq:ELBO_final}
	\end{equation} 
	where
	\begin{align}
		\tilde{p}_0^{(i)} &= \eta_{p_0 \vert p_t}^{(i)} + L_{p_0 \vert p_t}^{(i)^\mathrm{T}} \varepsilon\\
		(\eta_{p_0 \vert p_t}^{(i)}, \Gamma_{p_0 \vert p_t}^{(i)}) &= \Lambda(p_t^{(i)})  \\
		\varepsilon &\sim \mathcal{N}(0,I_d) 
	\end{align}
	where $\tilde{p}_0^{(i)}$ is a sample of the approximate posterior distribution $\pi_\phi(p_0 \vert p_t^{(i)})$ with mean $\eta_{p_0 \vert p_t}^{(i)}$ and covariance $\Gamma_{p_0 \vert p_t}^{(i)}$ given by the neural network $\Lambda$ and $L_{p_0 \vert p_t}^{(i)}$ is the Cholesky decomposition of the approximate posterior covariance matrix $\Gamma_{p_0 \vert p_t}^{(i)} = L_{p_0 \vert p_t}^{(i)^\mathrm{T}}L_{p_0 \vert p_t}^{(i)}$. Further, $\varepsilon$ is a draw from the standard normal distribution and $I_d$ is an identity matrix. 
	
	Now, since the posterior distribution is Gaussian, the first term in \eqref{eq:ELBO_final} is a KLD between two Gaussian distributions. Furthermore, by utilizing the linear observation model \eqref{eq:forwMod}, the maximization of the ELBO can be written as a minimization problem  
	\begin{align}
		\begin{split}
			\argmin_{w_\Lambda} \frac{1}{I} &\sum_{i = 1}^{I} 
			\Big(
			\mathrm{tr} \left( \Gamma_{p_0}^{-1} \Gamma_{p_0 \vert p_t}^{(i)} \right) + 
			\left\Vert L_{p_0} \left( \eta_{p_0 \vert p_t}^{(i)} - \eta_{p_0} \right) \right\Vert^2_2 +
			\log \frac{\left\vert \Gamma_{p_0} \right\vert}{\left\vert \Gamma_{p_0 \vert p_t}^{(i)} \right\vert} \\ &+  
			\left\Vert  L_{e^{(i)}} \left( p_t^{(i)} - K\tilde{p}_0^{(i)} - \eta_e\right) \right\Vert^2_2 \Big),
			\label{eq:VAE_min}
		\end{split}
	\end{align} 
	where
	\begin{align}
		\tilde{p}_0^{(i)} &= \eta_{p_0 \vert p_t}^{(i)} + L_{p_0 \vert p_t}^{(i)^\mathrm{T}} \varepsilon\\
		\Lambda(p_t^{(i)}) &= (\eta_{p_0 \vert p_t}^{(i)}, \Gamma_{p_0 \vert p_t}^{(i)}) \\
		\varepsilon &\sim \mathcal{N}(0,I_d),
	\end{align}
	and $w_\Lambda$ are the trainable parameters of the neural network $\Lambda$, and tr denotes the matrix trace. In the VAE, \eqref{eq:VAE_min} is regarded as a loss function that is minimized based on the dataset $P_t^{\textsc{VAE}} = \{p_t^{(i)}, \sigma_e^{(i)}\}_{i=1\dots,I}$.
	
	\subsubsection{Training}
	During the training process, the measurement data samples $p_t^{(i)}$ act as inputs to the neural network. On the other hand, the standard deviations of the measurement noise $\sigma_e^{(i)}$ are used to compute $L_e^{(i)}$ in the data likelihood term of the loss function \eqref{eq:VAE_min}. The training procedure of the VAE framework is summarized in Algorithm \eqref{alg:VAE}.
	\begin{algorithm}[h]
		\caption{Training the VAE}
		\label{alg:buildtree}
		\begin{algorithmic}
			\REQUIRE Training data $P_t^{\textsc{VAE}}$, prior mean $\eta_{p_0}$ and covariance $\Gamma_{p_0}$, forward operator $K$
			\ENSURE Neural network parameters $w_\Lambda$
			\REPEAT 
			\FOR{$\{p_t^{(i)}, \sigma_e^{(i)}\}_{i=1\dots,I} \in P_t^{\textsc{VAE}}$ }
			\STATE{Propagate data through the encoder: $(\eta_{p_0 \vert p_t}^{(i)}, \Gamma_{p_0 \vert p_t}^{(i)}) \leftarrow \Lambda(p_t^{(i)})$}
			\STATE{Sample standard normal distribution: $\varepsilon \leftarrow$ sample $\mathcal{N}(0,I_d)$}
			\STATE{Sample the approximate posterior: $\tilde{p}_0^{(i)} \leftarrow \eta_{p_0 \vert p_t}^{(i)} + L_{p_0 \vert p_t}^{(i)^\mathrm{T}} \varepsilon$}
			\STATE{Compute loss \eqref{eq:VAE_min}}
			\STATE{Update $w_\Lambda$}
			\ENDFOR
			\UNTIL{Stopping criterion is met}
			\RETURN $w_\Lambda$
		\end{algorithmic}
		\label{alg:VAE}
	\end{algorithm}
	
	Conventionally, the VAE architecture consists of two neural networks referred to as the encoder and the decoder \cite{Kingma2014}. In addition to the approximate posterior model $\pi_\phi(p_0 \vert p_t)$, the decoder representing the likelihood distribution $\pi_\theta(p_t \vert p_0)$ is parameterized by a set of learned parameters $\theta$. These parameters are then optimized jointly during the training process. In the proposed method, however, the role of the decoder is replaced by the data likelihood function and the training is performed solely on the encoder network. Utilizing this approach, the VAE, that was originally proposed as a generative  modeling framework, can be utilized in solving the Bayesian inverse problem.
	
	\subsubsection{Approximating the posterior distribution}
	Once the network parameters $w_\Lambda$ have been optimized, the statistics to the approximate posterior distribution are given by evaluating the network with a set of photoacoustic data
	\begin{equation}
		\Lambda(p_t) = (\eta_{p_0 \vert p_t}, \Gamma_{p_0 \vert p_t}).
		\label{eq:appr_evaluation}
	\end{equation}	
	In this work, the posterior distribution estimated by the VAE is modeled as an uncorrelated Gaussian distribution, leading to a diagonal covariance matrix.

	\subsection{Uncertainty quantification variational autoencoder}
	\label{subsec:UQ_VAE}
	
	The VAE presented in Section \ref{subsec:Auto-Encoding Variational Bayes} enables solving the variational Bayesian inverse problem using a neural network. The VAE, however, utilizes a training dataset consisting solely of samples of measurement data $p_t^{(i)}$ and the corresponding noise levels $\sigma_e^{(i)}$. It could, therefore, be beneficial to include additional information provided by the true initial pressure images corresponding to the data samples in the training procedure.
	
	Recently, it was proposed that the VAE could be extended by utilizing a family of Jensen-Shannon divergences (JSD) \cite{Goh2021}. Following the approach, the JSD between the approximate posterior distribution $\pi_\phi(p_0 \vert p_t)$ and the true posterior distribution $\hat{\pi}(p_0 \vert p_t)$ can be written as
	\begin{multline}
		\mathrm{JS}_\lambda(\pi_\phi(p_0 \vert p_t) \Vert \hat{\pi}(p_0 \vert p_t)) = 
		\lambda \mathrm{KL}(\pi_\phi(p_0 \vert p_t) \Vert (1-\lambda)\pi_\phi(p_0 \vert p_t) + \lambda \hat{\pi}(p_0 \vert p_t)) \label{eq:JSD} \\ 
		+ (1-\lambda)\mathrm{KL}(\hat{\pi}(p_0 \vert p_t) \Vert (1-\lambda)\pi_\phi(p_0 \vert p_t) + \lambda \hat{\pi}(p_0 \vert p_t)),
	\end{multline}
	where $\lambda \in\, ]0,1[$. Now, following the approach presented in \cite{Goh2021}, it can be shown that minimization of
	\begin{equation}
		\frac{1-\lambda}{\lambda}\mathrm{KL}(\hat{\pi}(p_0 \vert p_t) \Vert \pi_\phi(p_0 \vert p_t)) - \mathbb{E}_{p_0 \sim \pi_\phi(p_0 \vert p_t)} \left[ \log(\pi(p_t \vert p_0)) \right]
		+ \mathrm{KL}\left( \pi_\phi(p_0 \vert p_t) \Vert \pi(p_0) \right)
		\label{eq:Minimize_JS_KLD}
	\end{equation}
	could minimize a scaled sum of the JSD and the KLD
	\begin{equation}
		\frac{1}{\lambda}\mathrm{JS}_\lambda (\pi_\phi(p_0 \vert p_t) \Vert \hat{\pi}(p_0 \vert p_t)) +  \mathrm{KL} (\pi_\phi(p_0 \vert p_t) \Vert \hat{\pi}(p_0 \vert p_t)).
		\label{eq:Minimized_JS_KLD}
	\end{equation}
	resulting in a solution of the variational Bayesian inverse problem.
	
	Let us now consider minimization of the first term in \eqref{eq:Minimize_JS_KLD}. Let us introduce a training dataset consisting of initial pressure images $p_0^{(i)}$, the measurement data and measurement noise such that $P_t^{\textsc{UQ}} = \{p_t^{(i)}, p_0^{(i)}, \sigma_e^{(i)} \}, i= 1, \dots, I$. For minimization of the KLD between the true and the approximate posterior distributions, the authors in \cite{Goh2021} propose to approximate it as follows. First, minimization of a KLD is equivalent to maximization of the likelihood function with respect to $\phi$. Second, a Monte Carlo estimation of the likelihood function is formed by using the initial pressure data $p_0^{(i)}$. Third, the approximate posterior distribution is modeled as a Gaussian distribution. Using these assumptions, the first term in \eqref{eq:Minimize_JS_KLD} can be written as \cite{Goh2021}
	\begin{align}
		\begin{split}
			\mathrm{KL}(\hat{\pi}(p_0 \vert p_t) \Vert \pi_\phi(p_0 \vert p_t)) 
			&= \mathbb{E}_{p_0 \sim \pi_\phi(p_0 \vert p_t^{(i)})} \left[ - \log \left( \pi_\phi(p_0 \vert p_t^{(i)}) \right) \right] \\
			&\lesssim \mathbb{E}_{p_0 \sim \pi(p_0)}\left[ -\log\left( \pi_\phi(p_0 \vert p_t^{(i)}) \right) \right] \\
			&\approx -\log\left( \pi_\phi(p_0^{(i)}, p_t^{(i)}) \right) \\
			&= \frac{D}{2} \log(2\pi) + \frac{1}{2} \log \Big\vert \Gamma_{p_0 \vert p_t}^{(i)} \Big\vert + \frac{1}{2} \left\Vert L_{p_0 \vert p_t} \left( \eta_{p_0\vert p_t}^{(i)} - p_0^{(i)} \right) \right\Vert^2_2.
			\label{eq:UQVAE_KLD_term}
		\end{split}
	\end{align}
	
	Let us now model the initial pressure $p_0$ and noise $e$ as mutually independent and Gaussian distributed. Using these assumptions, the second and last terms in \eqref{eq:Minimize_JS_KLD} correspond to the minimization to the ELBO \eqref{eq:VAE_min}. Let us further consider a neural network $\Lambda$ that takes a set of measurement data $p_t^{(i)}$ as an input and outputs the statistics $\phi$ of the Gaussian approximate posterior distribution $\pi_\phi(p_0 \vert p_t)$. Then, by applying the reparameterization trick \cite{Kingma2014} and utilizing the linear observation model \eqref{eq:forwMod}, the problem of minimizing the JSD \eqref{eq:JSD} can be written as \cite{Goh2021} 
	\begin{align}
		\begin{split}
			\argmin_{w_\Lambda} \frac{1}{I} \sum_{i = 1}^{I} \frac{1-\lambda}{\lambda} 
			&\left( \log \left\vert \Gamma_{p_0 \vert p_t}^{(i)} \right\vert 
			+ \left\Vert L_{p_0 \vert p_t}^{(i)} \left( \eta_{p_0 \vert p_t}^{(i)} - p_0^{(i)} \right) \right\Vert ^2_2  \right) \\
			&+ \left\Vert L_{e^{(i)}} \left( p_t^{(i)} - K\tilde{p}_0^{(i)} - \eta_e \right) \right\Vert^2_2 \\
			&+ \mathrm{tr} \left( \Gamma_{p_0}^{-1} \Gamma_{p_0 \vert p_t}^{(i)} \right) + 
			\left\Vert L_{p_0} \left( \eta_{p_0 \vert p_t}^{(i)} - \eta_{p_0} \right) \right\Vert^2_2
			+ \log \frac{\left\vert \Gamma_{p_0} \right\vert}{\left\vert \Gamma_{p_0 \vert p_t}^{(i)} \right\vert},
			\label{eq:VAE_optimization}
		\end{split}
	\end{align}
	and $w_\Lambda$ denotes the trainable parameters of the neural network and $\tilde{p}_0^{(i)}, \eta_{p_0 \vert p_t}^{(i)}, \textrm{and}\, \Gamma_{p_0 \vert p_t}^{(i)}$ are as in \eqref{eq:VAE_min}.
	
	Compared to the VAE \eqref{eq:VAE_min}, the minimized functional of the UQ-VAE \eqref{eq:VAE_optimization} includes an additional term containing the initial pressure images $p_0^{(i)}$. The effect of this term to the minimization is adjusted using the scaling parameter $\lambda$. When $\lambda = 1$, \eqref{eq:VAE_optimization} reduces to the optimization problem of the VAE \eqref{eq:VAE_min}. 
	
	\subsubsection{Training}
	During the training process of the UQ-VAE, the measurement data samples $p_t^{(i)}$ act as inputs to the neural network. However, compared to the VAE, both the initial pressure images $p_0^{(i)}$ and the standard deviations of the measurement noise $\sigma_e^{(i)}$ are used to compute the value of the loss function \eqref{eq:VAE_optimization}. The training procedure of the UQ-VAE framework is summarized in Algorithm \eqref{alg:UQ-VAE}.
	\begin{algorithm}
		\caption{Training the UQ-VAE}
		\begin{algorithmic}
			\REQUIRE Training data $P_t^{\textsc{UQ}}$, prior mean $\eta_{p_0}$ and covariance $\Gamma_{p_0}$, forward, operator $K$, scaling parameter $\lambda$
			\ENSURE Neural network parameters $w_\Lambda$
			\REPEAT 
			\FOR{$\{p_t^{(i)}, p_0^{(i)}, \sigma_e^{(i)}\}_{i=1\dots,I} \in P_t^{\textsc{UQ}}$ }
			\STATE{Propagate data through the encoder:
				$(\eta_{p_0 \vert p_t}^{(i)}, \Gamma_{p_0 \vert p_t}^{(i)}) \leftarrow \Lambda(p_t^{(i)})$}
			\STATE{Sample standard normal distribution: $\varepsilon \leftarrow$ sample $\mathcal{N}(0,I_d)$}
			\STATE{Sample the approximate posterior: $\tilde{p}_0^{(i)} \leftarrow \eta_{p_0 \vert p_t}^{(i)} + L_{p_0 \vert p_t}^{(i)^\mathrm{T}} \varepsilon$}
			\STATE{Compute loss \eqref{eq:VAE_optimization}}
			\STATE{Update $w_\Lambda$}
			\ENDFOR
			\UNTIL{stopping criterion is met}
			\RETURN $w_\Lambda$
		\end{algorithmic}
		\label{alg:UQ-VAE}
	\end{algorithm}
	
	\subsubsection{Approximating the posterior distribution}
	Once the network parameters for the UQ-VAE have been optimized, the statistics to the approximate posterior distribution are given similarly as in the case of the VAE \eqref{eq:appr_evaluation}, by evaluating the network with a set of photoacoustic data. In this work, the posterior distribution estimated by the UQ-VAE is modeled as an uncorrelated Gaussian distribution with a diagonal covariance matrix.
	
	\section{Simulation setup}
	\label{sec:Simulations}
	
	In this work, the UQ-VAE framework was studied in the inverse problem of PAT. Performance of the approach was evaluated using multiple levels of noise and varying scaling parameter $\lambda$. Two sensor geometries were considered. The results were compared to the solution of the inverse problem in a Bayesian framework.
	
	\subsection{Simulation geometries, discretizations, and parameters}
	\label{subsec:Simulation geometries}
	
	In the simulations, a 2D 10 mm $\times$ 10 mm domain was considered. Two sensor geometries consisting of 64 sensors placed either on one side of the domain or two adjacent sides (32 sensors for each side). These sensor geometries are referred to as one side and two side sensor geometries, respectively. The simulated targets consisted of a set of blood vessel mimicking phantoms and a Shepp-Logan phantom that are described in more detail in Section \ref{subsec:Data simulation}. 
	
	Two different spatial and temporal discretizations were used for data simulation and inverse problems. For data simulation, the domain was discretized using a 230 $\times$ 230 pixel discretization. The temporal discretization was chosen based on a Courant-Friedrichs-Lewy number of 0.3. For the inverse problem, the domain was discretized in 128 $\times$ 128 pixels and temporal discretization was chosen such that $\Delta t \leq \Delta h / c$, where $\Delta h$ is the pixel size. The discretization parameters are summarized in Table \ref{tbl:discretizations}. 
	
	Prior model used in this work was the Gaussian Ornstein-Uhlenbeck (OU) distribution 
	\begin{equation}
		\Gamma_{p_0, ij} = \sigma_{p_0}^2 \exp \left\{ \frac{\Vert r_i - r_j \Vert}{l} \right\},
		\label{eq:OUPrior}
	\end{equation}
	where $\sigma_{p_0}$ is the standard deviation, $r_{ij}$ are pixel locations and $l$ is the characteristic length controlling the spatial correlation between pixels \cite{RasmussenBook}. The OU covariance belongs to the class of Mat{\'e}rn covariances, and has previously been found to be efficient and versatile for multiple types of imaged targets in PAT \cite{Sahlstrom2020,Tick2016,Tick2018}.
	
	\begin{table}[tb!]
		{\footnotesize
			\caption{Spatial (grid size $N_h$ and pixel side $\Delta h$) and temporal discretizations (time points $N_t$ and time step $\Delta t$) used in data simulation and in the solution of the inverse problem.}
			\begin{center}
				\begin{tabular}{ccccc} \hline \rule{0pt}{2.3ex} 
					& \bf $N_h$ & \bf $\Delta h \, (\mu m$) & \bf $N_t$ & $\Delta t \, (\mathrm{ns})$ \\ 
					\hline \rule{0pt}{2.3ex} 
					Simulation & 230$\times$230 & $ 43.5 $ & 1094 & 8.7 \\
					Inverse problem    & 128$\times$128 & $ 78.1$ & 512 & 18.6 \\ \hline
				\end{tabular}
			\end{center}
		}
		\label{tbl:discretizations}
	\end{table}

	\subsection{Data simulation}
	\label{subsec:Data simulation}
	
	Datasets for training and evaluating the performance of the neural network were created using 45 segmented retinal images from the High-Resolution Fundus Image Database \cite{Budai2013}. The images were first split into 43 training images and 2 testing images and rotated in 10$^\circ$ increments. Then, from each of the rotated images, sub-images were cropped using a sliding 600$\times$600 pixel window. These images were further interpolated to the 230$\times$230 pixel simulation grid (Table \ref{tbl:discretizations}) and scaled to have the maximum value of one. Then, to simulate more realistic targets, the zero-valued backgrounds of the images were replaced by draws of the OU prior \eqref{eq:OUPrior} with prior parameters $\sigma_{p_0} = 0.0625$, $\eta_{p_0} = 0$, and $l = 0.0005$. Furthermore, variability to the vessel amplitudes was created as follows. First, samples from the of the OU prior distribution with prior parameters $\sigma_{p_0} = 0.25$, $\eta_{p_0} = 0.05$, and $l = 0.005$ were drawn. Then, the pixels corresponding to the vessel areas in the images were indexed, and the corresponding pixels from the prior draws were subtracted from the vessel amplitudes. For all prior draws used to modify the phantoms, possible negative values were replaced by absolute values of the negative values. In total, 50000 training images and 2500 testing images were created. To further evaluate the ability of the neural network to generalize outside the training dataset, a Shepp-Logan phantom was used. The Shepp-Logan phantom was constructed in the 230 $\times$ 230 pixel simulation grid using an inbuilt MATLAB function. 
	
	Photoacoustic data corresponding to the phantoms in the one and two side geometries was simulated using the wave-equation \eqref{eq:Wave_equation} that was solved with the $k$-Wave MATLAB toolbox \cite{Treeby2010kwave}. Zero mean uncorrelated Gaussian noise was added to the training and testing datasets. The standard deviation of the noise was chosen as a percentage of the maximum simulated amplitude for each dataset. For the training dataset, the percentages for each simulated data were chosen randomly from between 1\% and 5\% (corresponding to approximate signal to noise ratio interval of 22dB to 8dB). For the vessel testing datasets, three constant noiselevels of 1\%, 3\% and 5\% were considered. For the Shepp-Logan phantom, noise level of 3\% was used. Finally, in order to simulate a more realistic training procedure and to avoid the inverse crime, both the phantoms and the corresponding simulated data in the training and testing datasets were interpolated to the inverse problem discretization (Table \ref{tbl:discretizations}). The training, validation and test datasets are summarised in Table \ref{tbl:datasets}.
	
	\begin{table}[tb!]
		{\footnotesize
			\caption{Training, validation, and testing datasets used in the simulations.}
			\begin{center}
				\begin{tabular}{cccc} \hline\rule{0pt}{2.3ex}
					Dataset & Phantom type & Samples & Noise level  \\ \hline \rule{0pt}{2.3ex}
					Training & Vessel & 45000 & Variable 1\%-5\% \\
					Validation & Vessel & 5000 & Variable 1\%-5\%\\ 
					Testing 1 & Vessel & 2500 & Constant 1\%, 3\%, 5\%\\
					Testing 2 & Shepp-Logan & 1 & Constant 3\%\\ \hline
				\end{tabular}
			\end{center}
		}
		\label{tbl:datasets}
	\end{table}

	\subsection{Inverse problem}
	\label{subsec:Inverse problem}
	
	The spatial and temporal discretizations used in the inverse problem are shown in Table \ref{tbl:discretizations}. The prior distribution used in the inverse problem was the OU prior distribution \eqref{eq:OUPrior} with mean $\eta_{p_0} = 0.5$, standard deviation $\sigma_{p_0} = 0.25$ and characteristic length $l = 0.5$ mm. The expected value of the prior was chosen as the mean value between the minimum and maximum values of the initial pressures. Furthermore, the standard deviation of the prior was chosen such that 95.5\% of the initial pressure values lied within $\pm2$ standard deviations from the mean value. The characteristic length was chosen as 5\% of the domain side length in order to allow for sharper changes in the reconstructions. The noise was modelled using the statistics of the noise added to the simulated data. 
	
	The inverse problem was solved using the UQ-VAE \eqref{eq:VAE_optimization} with scaling parameter $\lambda$ values $\lambda$ = 1, 0.5, 0.1, 0.05. Parameter choice $\lambda = 1$ corresponds to the solution of the VAE \eqref{eq:VAE_min}. To reduce the size of the neural networks, the posterior distribution in the UQ-VAE was modeled as uncorrelated Gaussian distribution leading to a diagonal covariance matrix. Implementation of the neural network is presented in Section \ref{subsec:VAE and UQ-VAE}. The results were compared against the solution of the inverse problem in a Bayesian framework \eqref{eq:Posterior_exp}-\eqref{eq:Posterior_mean}. 
	
	\subsection{UQ-VAE and VAE}
	\label{subsec:VAE and UQ-VAE}
	
	\begin{figure}[tbp!]
		\centering
		\includegraphics[width=\textwidth]{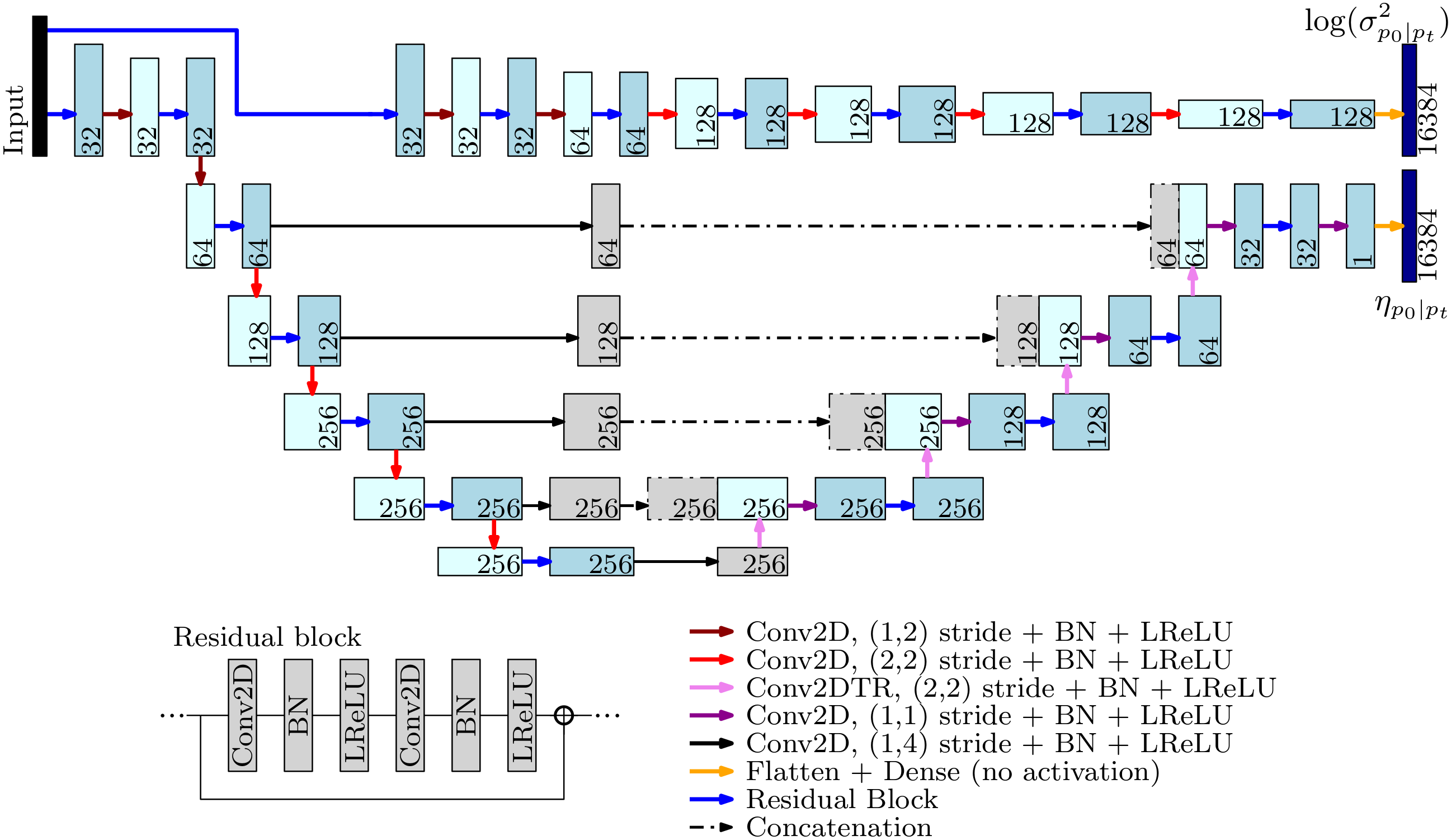}
		\caption{Neural network architecture. The neural network consists of two branches outputting the expected value $\eta_{p_0 \vert p_t}$ and logarithm of variances $\log(\sigma^2_{p_0\vert p_t})$ to the approximate posterior distribution $\pi_\phi(p_0 \vert p_t)$.  In the figure, the numbers denote the number of filters for convolutional layers or the number of nodes in densely connected layers. For the convlution layers, 3$\times$3 pixel convolution kernels were used in the mean branch and 5$\times$5 pixel convolution kernels in the standard deviation branch.}
		\label{fig:NN_arc}
	\end{figure}
	
	The neural network $\Lambda$ was constructed with separate branches for estimating the mean and standard deviation of the approximate posterior distribution. The network architecture is shown in Fig. \ref{fig:NN_arc}. The branch for the mean was constructed by combining an asymmetric U-Net with a densely connected output layer. The branch for the standard deviations consisted of a contracting convolutional neural network with a densely connected output layer. In both branches, the convolutional layers were replaced by residual blocks to improve the learning capability of the network. Addition of the densely connected output layers was found to result in more data-consistent results via smaller data likelihood values and subsequently improved reactivity to measurement noise present in the input data. The two branch structure of the neural network was used to reduce the memory requirements as the U-Net with two densely connected output layers could not be used within the available memory resources. Furthermore, the simpler structure of the standard deviation branch was motivated by the simpler structure of the standard deviation images compared to the mean images.
	
	The neural networks for the UQ-VAE and VAE were trained as outlined in Algorithms \eqref{alg:VAE} and \eqref{alg:UQ-VAE} by minimizing the functionals \eqref{eq:VAE_min} and \eqref{eq:VAE_optimization}. Training was performed separately for each sensor geometry and scaling parameter $\lambda$. Once the networks were trained, the posterior distributions were estimated using a photoacoustic dataset as an input to the network.
	
	The neural networks were implemented in Python 3.7.9 and Tensorflow Keras 2.1.0. The networks were trained using the Adam optimizer with a batch size of 10. The learning rate was set as decaying polynomial learning rate with an initial learning rate of $50 \cdot 10^{-6}$ and final learning rate of $5 \cdot 10^{-6}$ after 30 epochs after which the learning rate stayed constant. The training was performed using an NVIDIA GeForce GTX Titan GPU with 12GB of VRAM. For training of the neural network, the training dataset was split into 45000 training and 5000 validation samples.

	\section{Results}
	\label{sec:Results}
	
	\subsection{Vessel phantoms}
	
	\subsubsection{Varying scaling parameter}
	
	\begin{figure}[!tbp]
		\centering
		\includegraphics[]{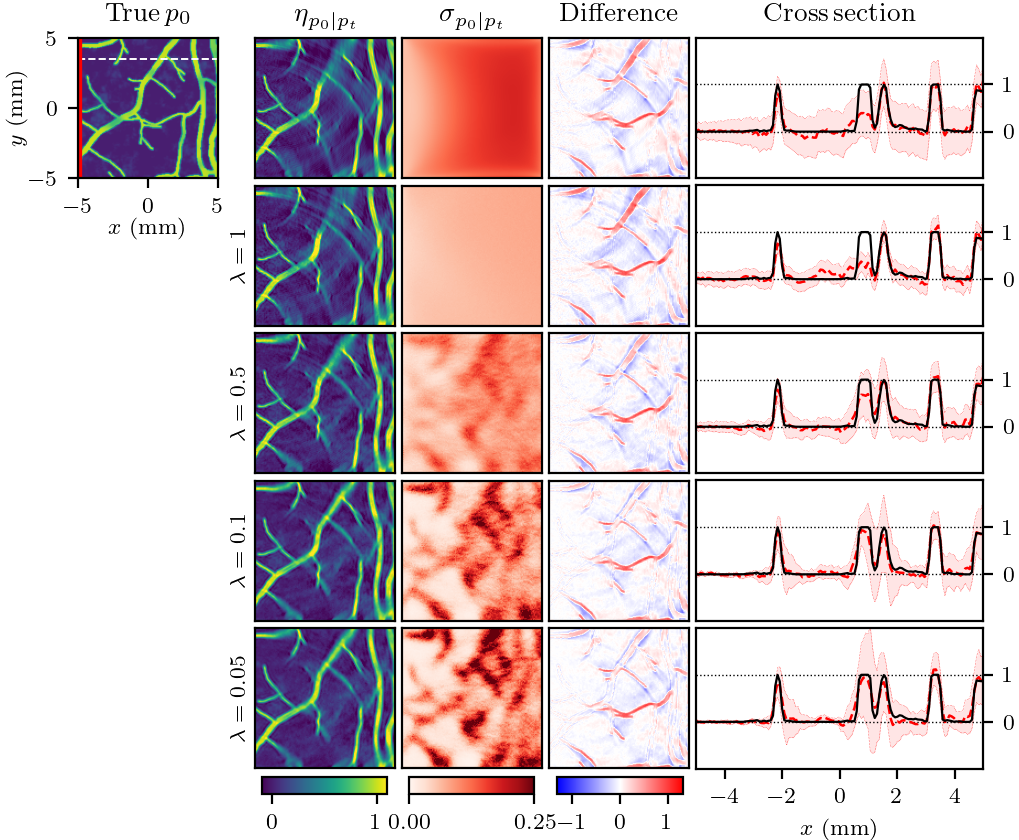}
		\caption{Estimated posterior distribution in a one side sensor geometry with 3\% noise level. Columns from left to right. The true initial pressure $p_0$ image (first column), the estimated posterior mean $\eta_{p_0 \vert p_t}$ (second column) and standard deviation $\sigma_{p_0\vert p_t}$ (third column), difference between the true initial pressure and the estimated mean (fourth column), and a cross section of the mean through the target with $\pm$3 sd credibility interval (fifth column). Images from top to bottom: Bayesian approach (first row), UQ-VAE with scaling parameter values $\lambda$ = 1 (VAE) (second row) and $\lambda$ = 0.5, 0.1, and 0.05 (rows 3-5). The location of the sensors is indicated with a solid red line and the location of the cross-section is indicated with a white dashed line in the first column image. }
		\label{fig:LAMBDA_VESSEL_OS}
	\end{figure}
	\begin{figure}[!tbp]
		\centering
		\includegraphics[]{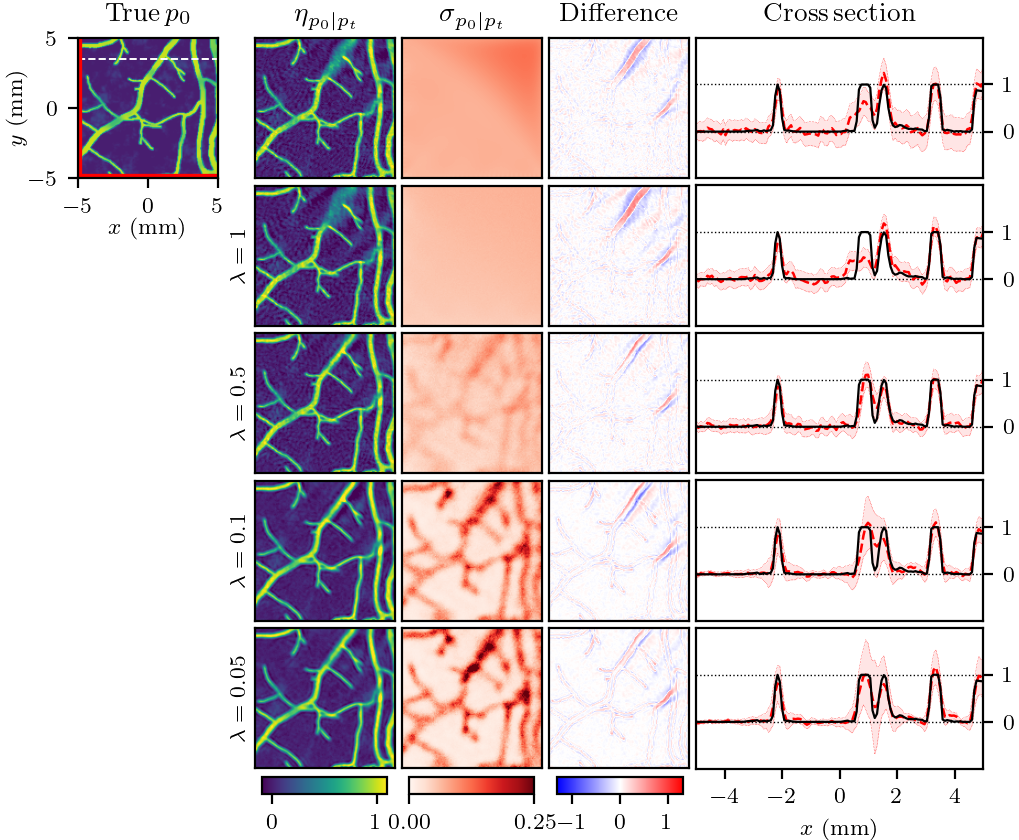}
		\caption{Estimated posterior distribution in a two side sensor geometry with 3\% noise level. Columns from left to right. The true initial pressure $p_0$ image (first column), the estimated posterior mean $\eta_{p_0 \vert p_t}$ (second column) and standard deviation $\sigma_{p_0\vert p_t}$ (third column), difference between the true initial pressure and the estimated mean (fourth column), and a cross section of the mean through the target with $\pm$3 sd credibility interval (fifth column). Images from top to bottom: Bayesian approach (first row), UQ-VAE with scaling parameter values $\lambda$ = 1 (VAE) (second row) and $\lambda$ = 0.5, 0.1, and 0.05 (rows 3-5). The location of the sensors is indicated with a solid red line and the location of the cross-section is indicated with a white dashed line in the first column image.  }
		\label{fig:LAMBDA_VESSEL_L}
	\end{figure}
	Posterior mean and standard deviations estimated using the UQ-VAE with scaling parameters $\lambda=1$ (VAE) and $\lambda$ = 0.5, 0.1, 0.05 (UQ-VAE) are shown in Figs. \ref{fig:LAMBDA_VESSEL_OS} and \ref{fig:LAMBDA_VESSEL_L} for the one and two side sensor geometries, respectively. The figures also show the posterior mean and standard deviations estimated using the Bayesian approach \eqref{eq:Posterior_mean}-\eqref{eq:Posterior_cov}. The noise level was 3\%.  
	
	As it can be seen, the expected values given by the UQ-VAE vary depending on the choice of the scaling parameter $\lambda$. As $\lambda \rightarrow 1$, the expected values start to resemble the conventional Bayesian solution. On the other hand, as $\lambda \rightarrow 0$, the blurring due to the limited view sensor geometry is partially corrected resulting in a reconstruction closer to the simulated ground truth image. In the case of the two side sensor geometry and small values of $\lambda$, it can also be seen that the background noise in the images is alleviated. However, in these cases the sharp edges of the vessel structures are slightly smoothed.
	
	Regarding the standard deviations given by the UQ-VAE, it can be observed that the estimated standard deviations are lower closer to the sensors and in the region closed by the sensors. That is, the solution of the inverse problem can be regarded as more reliable in these regions. Additionally, the standard deviation values estimated by the UQ-VAE increase as $\lambda \rightarrow 0$. Furthermore, as $\lambda \rightarrow 0$, the higher magnitudes of the standard deviation concentrate in areas of higher expected values and the areas where the limited view artifacts are present. 
	
	The variability of the standard deviation values with respect to the scaling parameter $\lambda$ can also be observed in the cross sections and the corresponding credibility intervals. In the case of $\lambda = 1$, the true value of the initial pressure lies partially outside the $\pm$3 standard deviation credibility interval. In the case of smaller values of $\lambda$, compensation of the limited view artifacts and the larger values of standard deviations lead to the true values being mostly within the $\pm$3 standard deviation credibility intervals.

	\subsubsection{Varying noise levels}
	
\begin{figure}[!tbp]
		\centering
		\includegraphics[]{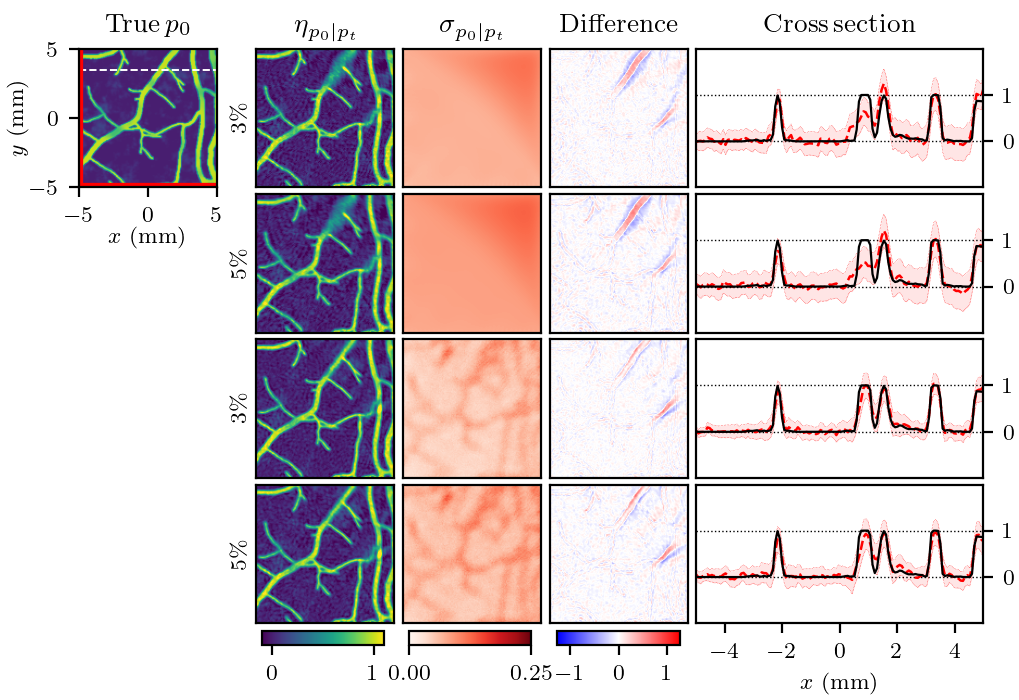}
		\caption{Estimated posterior distribution in a two side sensor geometry with 3\% and 5\% noise levels. Columns from left to right. The true initial pressure $p_0$ image (first column), the estimated posterior mean $\eta_{p_0 \vert p_t}$ (second column) and standard deviation $\sigma_{p_0\vert p_t}$ (third column), difference between the true initial pressure and the estimated mean (fourth column), and a cross section of the mean through the target with $\pm$3 sd credibility interval (fifth column). Images from top to bottom: Bayesian approach for noise levels 3\% and 5\% (rows 1-2), UQ-VAE with scaling parameter value $\lambda$ = 0.5 and noise levels 3\% and 5\% (rows 3-4). The location of the sensors is indicated with a solid red line and the location of the cross-section is indicated with a white dashed line in the first column image.}
		\label{fig:NOISE_VESSEL_L}
	\end{figure}

	Posterior mean and standard deviations estimated using the UQ-VAE with noise levels 3\% and 5\%, scaling parameter $\lambda=0.5$, and the two side sensor geometry are shown in Fig. \ref{fig:NOISE_VESSEL_L}. The figure also shows the posterior mean and standard deviations estimated using the Bayesian approach \eqref{eq:Posterior_mean}-\eqref{eq:Posterior_cov}. From the results it can be seen that the magnitude of the standard deviations for the conventional and the UQ-VAE increase with larger noise indicating that the neural network is able to react to the noise level present in the measurement data.
	
	\subsubsection{Relative errors and computation times}
	
	In addition to visual comparison, the results given by the VAE, the UQ-VAE, and the Bayesian approach were compared using relative error between the true initial pressure $p_0^{\textsc{true}}$ and the estimated initial pressure $p_0^{\textsc{est}}$
	\begin{equation}
		\mathrm{RE} = 100\% \frac{\Vert p_0^{\textsc{true}} - p_0^{\textsc{est}} \Vert}{\Vert p_0^{\textsc{true}} \Vert},
	\end{equation}
	structural similarity index (SSIM), peak signal to noise ratios (PSNR), and credibility percentages (CI). The credibility percentages were defined as the percentage of pixels for which the true value of the initial pressure lied within $\pm$3 standard deviation credibility interval (corresponding to 99.7\%) of the estimated posterior distribution. Furthermore, the effect of the scaling parameter $\lambda$ on the data likelihood was studied by computing the data fit values
	\begin{equation}
		\mathrm{DF} = \Vert p_t - Kp_0^{\textsc{est}} \Vert.
	\end{equation}
	Relative errors, structural similarity indices, credibility percentages and data fit values for the Bayesian approach, UQ-VAE with scaling parameter values $\lambda$ = 1, 0.5, 0.1, 0.05 and noise levels 1\%, 3\%, and 5\% for the one and two side sensor geometries are shown in Table \ref{tbl:Both_Side_Metrics}.
	\begin{table}[tbp!]
		\setlength{\tabcolsep}{3.5pt}
		{\footnotesize
			\caption{Relative errors (RE), structural similary indices (SSIM), peak signal to noise ratios (PSNR), and credibility percentages (CI), data fit values (DF) for the Bayesian approach and the UQ-VAE for scaling parameter values $\lambda$ = 1, 0.5, 0.1, and 0.05, noiselevels of 1\%, 3\%, and 5\%, and one and two side sensor geometries.}
			\begin{center}
				\begin{tabular}{p{1.5cm}cccp{0.1cm}ccc} 
					\multicolumn{8}{l}{\textsc{ONE SIDE}}\\
					\hline \\[-2ex]
					& \multicolumn{3}{c}{RE (\%)} &&\multicolumn{3}{c}{SSIM}\\
					\cline{2-4} \cline{6-8}
					\\[-2ex]
					Noiselevel    &1\%&3\%&5\%&&1\%&3\%&5\%\\
					\hline \\[-2ex]
					$\textrm{Bayes}$ &50.1$\pm$6.0& 49.6$\pm$6.5& 51.0$\pm$6.5&& 0.39$\pm$0.05& 0.41$\pm$0.05& 0.40$\pm$0.05\\
					$\lambda = 1$    &53.3$\pm$6.4& 52.8$\pm$6.6& 53.1$\pm$6.5&& 0.39$\pm$0.05& 0.39$\pm$0.05& 0.39$\pm$0.05\\ 
					$\lambda = 0.5$  &47.5$\pm$6.3& 47.5$\pm$6.4& 48.5$\pm$6.4&& 0.44$\pm$0.05& 0.44$\pm$0.05& 0.42$\pm$0.05\\
					$\lambda = 0.1$  &45.8$\pm$6.4& 46.3$\pm$6.4& 47.5$\pm$6.4&& 0.49$\pm$0.05& 0.48$\pm$0.05& 0.46$\pm$0.05\\
					$\lambda = 0.05$ &\textbf{45.6$\pm$6.5}& \textbf{46.2$\pm$6.4}& \textbf{47.4$\pm$6.4}&& \textbf{0.50$\pm$0.05}& \textbf{0.49$\pm$0.05}& \textbf{0.46$\pm$0.05}\\
					\\[-2ex]
					
					& \multicolumn{3}{c}{PSNR (dB)} &&\multicolumn{3}{c}{CI(\%)}\\
					\cline{2-4} \cline{6-8}
					\\[-2ex]
					Noiselevel    &1\%&3\%&5\%&&1\%&3\%&5\%\\
					\hline \\[-2ex]
					Bayes            &15.3$\pm$1.5& 15.4$\pm$1.6& 15.3$\pm$1.6&& 89.9$\pm$3.2& 93.4$\pm$2.7& 94.2$\pm$2.4\\
					$\lambda = 1$    &14.7$\pm$1.6& 14.8$\pm$1.6& 14.8$\pm$1.6&& 45.4$\pm$9.5& 83.5$\pm$5.9& 86.9$\pm$4.9\\ 
					$\lambda = 0.5$  &15.7$\pm$1.7& 15.8$\pm$1.7& 15.6$\pm$1.7&& 84.5$\pm$4.1& 94.3$\pm$1.9& 95.2$\pm$1.6\\
					$\lambda = 0.1$  &16.1$\pm$1.7& 16.0$\pm$1.7& 15.7$\pm$1.7&& 93.2$\pm$1.6& 96.3$\pm$0.9& 96.6$\pm$0.9\\
					$\lambda = 0.05$ &\textbf{16.1$\pm$1.7}& \textbf{16.0$\pm$1.7}& \textbf{15.8$\pm$1.7}&& \textbf{95.1$\pm$1.2}& \textbf{96.8$\pm$0.8}&\textbf{ 96.7$\pm$0.8}\\ 
					\\[-2ex]
					
					& \multicolumn{3}{c}{DF} &&\\
					\cline{2-4} 
					\\[-2ex]
					Noiselevel    &1\%&3\%&5\%&&&&\\
					\cline{1-4}  \\[-2ex]
					Bayes            &\textbf{1.23$\pm$0.21}&\textbf{3.85$\pm$0.66}&\textbf{6.88$\pm$1.23}&&&&\\
					$\lambda = 1$    &2.23$\pm$0.34&4.22$\pm$0.72&7.03$\pm$1.27&&&&\\ 
					$\lambda = 0.5$  &2.46$\pm$0.37&4.42$\pm$0.75&7.12$\pm$1.29&&&&\\
					$\lambda = 0.1$  &3.00$\pm$0.44&4.79$\pm$0.78&7.44$\pm$1.32&&&& \\
					$\lambda = 0.05$ &3.40$\pm$0.50&5.07$\pm$0.80&7.62$\pm$1.33&&&&\\ 
					\\[2ex]
					
					\multicolumn{8}{l}{\textsc{TWO SIDE}}\\
					\hline \\[-2ex]
					& \multicolumn{3}{c}{RE (\%)} &&\multicolumn{3}{c}{SSIM}\\
					\cline{2-4} \cline{6-8}
					\\[-2ex]
					Noiselevel    &1\%&3\%&5\%&&1\%&3\%&5\%\\
					\hline \\[-2ex]
					$\textrm{Bayes}$ & 24.8$\pm$3.1& 23.4$\pm$3.8& 26.1$\pm$4.0&& 0.62$\pm$0.04& 0.63$\pm$0.04& 0.60$\pm$0.05\\
					$\lambda = 1$    & 26.4$\pm$4.3& 26.3$\pm$4.3& 27.8$\pm$4.3&& 0.61$\pm$0.04& 0.61$\pm$0.04& 0.58$\pm$0.05\\ 
					$\lambda = 0.5$  & 21.5$\pm$3.5& \textbf{21.6$\pm$3.6}& \textbf{23.7$\pm$3.6}&& 0.68$\pm$0.04& 0.67$\pm$0.04& 0.63$\pm$0.04\\
					$\lambda = 0.1$  & \textbf{20.7$\pm$3.5}& 21.8$\pm$3.6& 23.8$\pm$3.6&& 0.77$\pm$0.03& 0.74$\pm$0.03& 0.69$\pm$0.03\\
					$\lambda = 0.05$ & 21.0$\pm$3.5& 22.3$\pm$3.6& 24.2$\pm$3.7&& \textbf{0.79$\pm$0.03}& \textbf{0.75$\pm$0.03}& \textbf{0.70$\pm$0.03}\\
					\\[-2ex]
					
					& \multicolumn{3}{c}{PSNR (dB)} &&\multicolumn{3}{c}{CI(\%)}\\
					\cline{2-4} \cline{6-8}
					\\[-2ex]
					Noiselevel    &1\%&3\%&5\%&&1\%&3\%&5\%\\
					\hline \\[-2ex]
					Bayes            &21.4$\pm$1.5& 21.9$\pm$1.7& 21.4$\pm$1.6&& 95.2$\pm$2.0& 97.4$\pm$1.5& 97.7$\pm$1.4\\
					$\lambda = 1$    &20.9$\pm$1.8& 20.9$\pm$1.8& 20.4$\pm$1.7&& 56.7$\pm$9.1& 94.9$\pm$2.3& 96.7$\pm$1.7\\ 
					$\lambda = 0.5$  &22.7$\pm$1.9& \textbf{22.6$\pm$1.8}& 21.8$\pm$1.7&& 86.1$\pm$5.1& 98.4$\pm$0.9& \textbf{98.8$\pm$0.8}\\
					$\lambda = 0.1$  &\textbf{23.0$\pm$1.9}& 22.5$\pm$1.8& \textbf{21.8$\pm$1.7}&& \textbf{97.0$\pm$0.9}& \textbf{98.6$\pm$0.5}& 98.7$\pm$0.4\\
					$\lambda = 0.05$ &22.8$\pm$1.8& 22.3$\pm$1.7& 21.6$\pm$1.6&& 96.8$\pm$0.8& 98.0$\pm$0.5& 98.0$\pm$0.5\\ 
					\\[-2ex]
					
					& \multicolumn{3}{c}{DF} &&\\
					\cline{2-4} 
					\\[-2ex]
					Noiselevel    &1\%&3\%&5\%&&&&\\
					\cline{1-4}  \\[-2ex]
					Bayes            &\textbf{1.19$\pm$0.17}&\textbf{3.84$\pm$0.57}&\textbf{7.03$\pm$1.07}&&&&\\
					$\lambda = 1$    &2.46$\pm$0.33&4.42$\pm$0.65&7.33$\pm$1.10&&&&\\ 
					$\lambda = 0.5$  &2.38$\pm$0.34&4.42$\pm$0.65&7.30$\pm$1.15&&&&\\
					$\lambda = 0.1$  &3.34$\pm$0.43&5.24$\pm$0.70&7.97$\pm$1.18&&&&\\
					$\lambda = 0.05$ &3.76$\pm$0.49&5.50$\pm$0.73&8.15$\pm$1.20&&&&\\ 
					
					\hline
					
				\end{tabular}
			\end{center}
			
		}
		\label{tbl:Both_Side_Metrics}
	\end{table}
	
	As it can be seen, the relative errors for the UQ-VAE are close to the relative errors of the conventional Bayesian approach. These values  decrease with smaller values of $\lambda$, indicating a slight increase in the accuracy of the estimates. In addition to the relative errors, the increase in the accuracy of the estimates is supported by the SSIM and PSNR values that increase with smaller values of $\lambda$. The credibility percentages for the UQ-VAE indicate that the standard deviations given by the VAE ($\lambda$ = 1) using small noise values are overly small leading to small credibility intervals and low credibility percentages. Furthermore, the credibility percentages increase with decreasing $\lambda$ due to increasing standard deviation values and increased accuracy of the estimated images. From the data fit values it can be observed, that the data fit values increase with increasing noise and decreasing values of $\lambda$. The data fit values for the UQ-VAE are, however, slightly higher compared to the Bayesian approach.
	
	Computation times for the posterior distribution using Bayesian approach and the UQ-VAE were 8.44 s and 0.12s, respectively. Training of the neural network was done separately for each scaling parameter $\lambda$ and sensor geometry. Depending on $\lambda$, convergence was achieved between 30-100 epochs. In general, smaller values of $\lambda$ resulted in faster convergence. Training time for one epoch was 34 minutes resulting in total training time between 17 and 57 hours.

	\subsection{Shepp-Logan phantom}
	
	\begin{figure}[!tbp]
		\centering
		\includegraphics[]{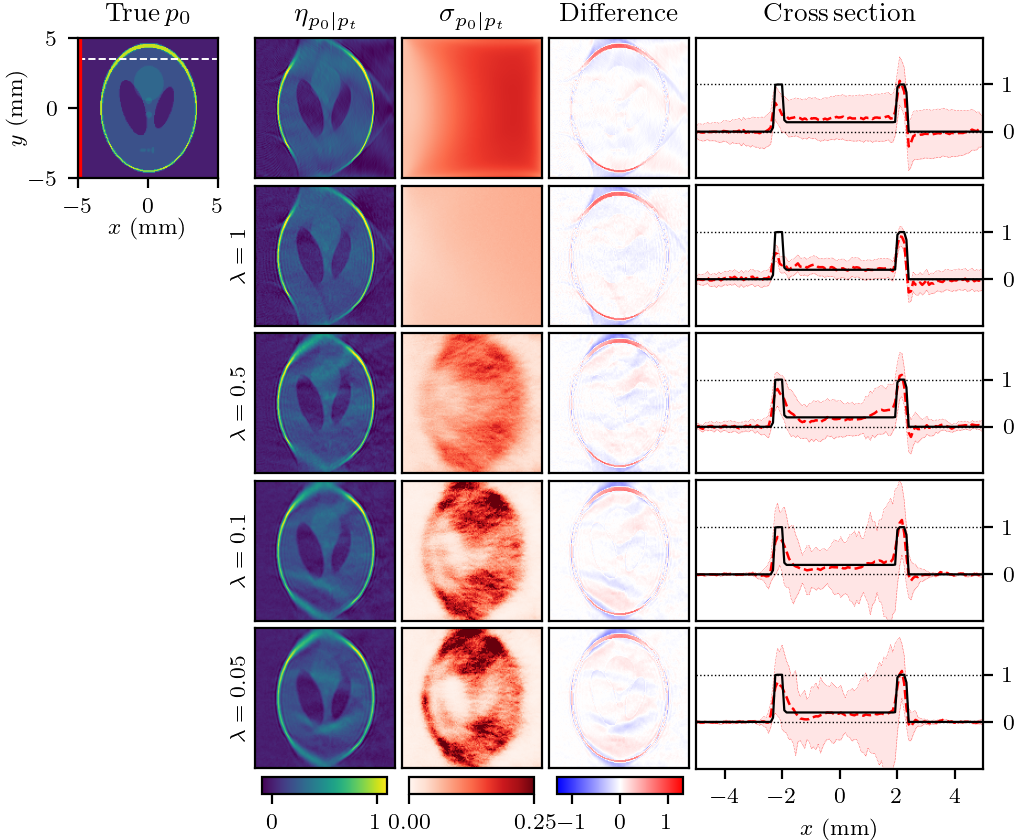}
		\caption{Estimated posterior distribution in a one side sensor geometry with 3\% noise level for the Shepp-Logan phantom. Columns from left to right. The true initial pressure $p_0$ image (first column), the estimated posterior mean $\eta_{p_0 \vert p_t}$ (second column) and standard deviation $\sigma_{p_0\vert p_t}$ (third column), difference between the true initial pressure and the estimated mean (fourth column), and a cross section of the mean through the target with $\pm$3 sd credibility interval (fifth column). Images from top to bottom: Bayesian approach (first row), UQ-VAE with scaling parameter values $\lambda$ = 1 (VAE) (second row) and $\lambda$ = 0.5, 0.1, and 0.05 (rows 3-5). The location of the sensors is indicated with a solid red line and the location of the cross-section is indicated with a white dashed line in the first column image.
		}
		\label{fig:LAMBDA_SHEPP_OS}
	\end{figure}
	
	\begin{figure}[!tbp]
		\centering
		\includegraphics[]{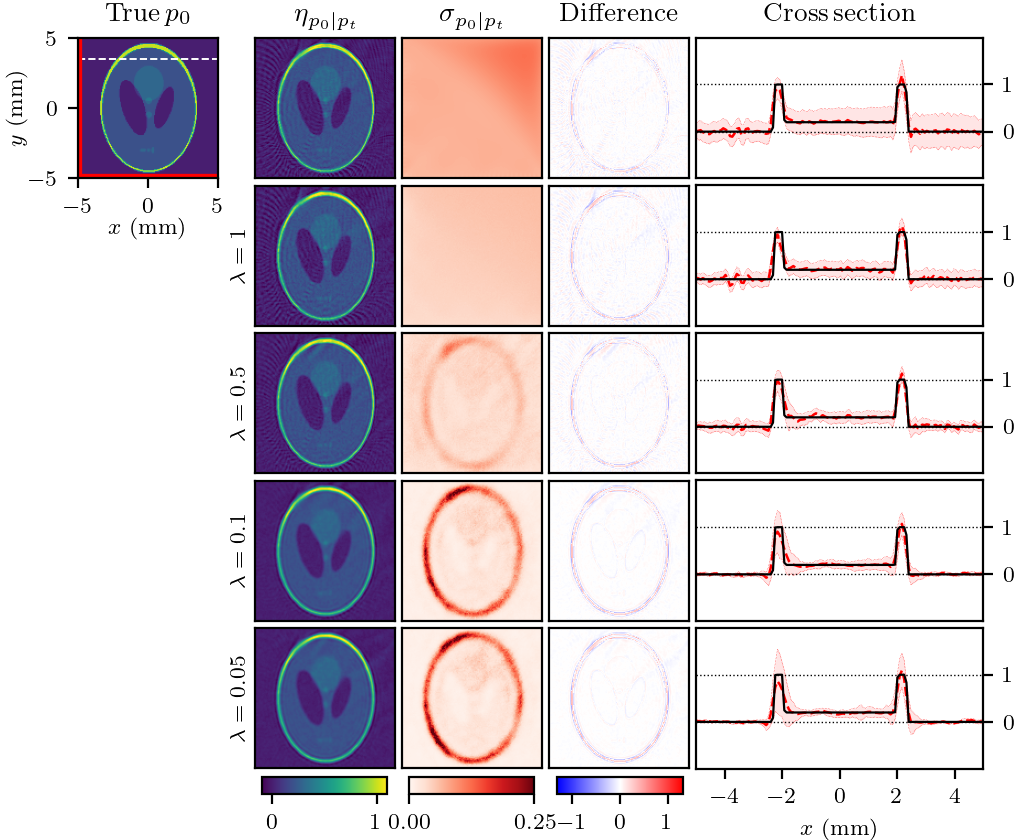}
		\caption{Estimated posterior distribution in a two side sensor geometry with 3\% noise level for the Shepp-Logan phantom. Columns from left to right. The true initial pressure $p_0$ image (first column), the estimated posterior mean $\eta_{p_0 \vert p_t}$ (second column) and standard deviation $\sigma_{p_0\vert p_t}$ (third column), difference between the true initial pressure and the estimated mean (fourth column), and a cross section of the mean through the target with $\pm$3 sd credibility interval (fifth column). Images from top to bottom: Bayesian approach (first row), UQ-VAE with scaling parameter values $\lambda$ = 1 (VAE) (second row) and $\lambda$ = 0.5, 0.1, and 0.05 (rows 3-5). The location of the sensors is indicated with a solid red line and the location of the cross-section is indicated with a white dashed line in the first column image.
		}
		\label{fig:LAMBDA_SHEPP_L}
	\end{figure}

	Posterior mean and standard deviations for the Shepp-Logan phantom estimated using the UQ-VAE with scaling parameters $\lambda=1$ (VAE) and $\lambda$ =0.5, 0.1, 0.05 (UQ-VAE) are shown in Figs. \ref{fig:LAMBDA_SHEPP_OS} and \ref{fig:LAMBDA_SHEPP_L} for the one and two side sensor geometries, respectively. The figures also show the posterior mean and standard deviations estimated using the Bayesian approach \eqref{eq:Posterior_mean}-\eqref{eq:Posterior_cov}. The noise level was 3\%.  
	
	As it can be seen, the expected values given by the UQ-VAE are close to the true initial pressure, indicating a good generalization ability of the neural network. In general, variability of the result given by the UQ-VAE with respect to the scaling parameter $\lambda$ can be explained similarly as in the case of the vessel phantoms. As $\lambda \rightarrow 1$, the expected value of the posterior distribution is close to the expected value given by the conventional approach and the magnitude of the standard deviation is smaller. Furthermore, as $\lambda \rightarrow 0$, the limited view artifacts are partially corrected, magnitudes of the standard deviation increase and localize in the areas of higher expected values.

	\section{Discussion}
	\label{sec:Discussion}
	
	The results given by the UQ-VAE framework were found to be highly dependent on the scaling parameter $\lambda$. In general, the effect of the scaling parameter $\lambda$ can be summarized as follows. Firstly, $\lambda$ acts as a scaling term for the so-called posterior term. As small values of $\lambda$ emphasize the minimization of the term containing the ground truth images, the reconstructions given by the network are closer to the ground truth images and exhibit less limited view artifacts. On the other hand, as $\lambda \rightarrow 1$, the data likelihood term is emphasized resulting reconstructions with more clear limited view artifacts. Secondly, the effect of $\lambda$ in the estimates of the posterior standard deviations can be observed both in the spatial locations and in the magnitudes. As $\lambda \rightarrow 0$, the magnitude of the uncertainty estimates increases and localizes in the areas of high expected values. Furthermore, as $\lambda \rightarrow 1$, the standard deviations are lower. That is, the choice of $\lambda$ affects both the accuracy of the estimates and their reliability, necessitating finding of suitable trade-off between the reconstruction characteristics and reasonable uncertainty estimates.  
	
	As with most deep learning frameworks, the choice of the neural network architecture plays a crucial role. On one hand, the architecture should be expressive enough to be able to learn the underlying problem. On the other hand, excessively large network structures can lead to unnecessarily long training times and memory requirements. In this work, choice of the network structure was motivated by the following points. Firstly the architecture comprising of two separate branches was used to reduce the memory requirements of the network. Secondly, the choice of standard deviation branch was motivated by the simpler structure of the standard deviation images that allowed for a simpler network architecture. Even though this network architecture was found to perform well within the hardware resources used in this work, the network architecture could be optimized further.
	
	An important factor to consider in the context of machine learning is the ability of the network to generalize to targets outside the training dataset. In this work it was found that the neural network was able perform well for the testing dataset containing vessel phantoms and the Shepp-Logan phantom. The good generalization ability of the neural network can be attributed to two factors, namely the inclusion of the measurement data likelihood term in the optimized functional and the use of an accurate forward operator during the training process. This enabled the neural network to learn from the information presented by the true forward operator.
	
	In this work, the approximate posterior distribution was modeled as uncorrelated Gaussian distribution, and thus only standard deviations of the posterior distribution were approximated. The method could, however, be extended for evaluating the full posterior covariance matrices with the expense of increasing computational cost. In this case, the computation cost could be reduced by estimating the Cholesky decomposition of the covariance matrix or by utilizing dimensionality reduction techniques such as the singular value decomposition.
	
	In this work, the Bayesian inverse problem of PAT was approached in a 2D setting. PAT is, however, inherently a high resolution 3D imaging modality. Extension of the proposed framework to 3D setting at its current form is challenging due to increasing network memory requirements as the number of unknowns increase. Therefore, methods for alleviating memory requirements, for example by model reduction, would be needed. Furthermore, the size and architecture of the neural network could also be optimized further by considering purely convolutional architectures. 
	
	Conventional implementations of the VAE consist of an encoder and a decoder. In this work, however, only the encoder was used and the data likelihood was computed using a forward operator. The current approach could be modified to resemble the conventional VAE by replacing the forward operator by a second neural network. This approach was studied in \cite{Goh2021} using a 2D steady state heat conduction problem. In that work, the forward operator was successfully learned, when sufficiently large amount of training data was available. Utilizing the learned forward operator could, therefore, be feasible in the case of PAT. Using this approach could, however, result in worse ability of the neural network to generalize especially with small amounts of training data.
	
	The major challenge of extending various neural network based approaches to experimental setting is the lack of large amounts of experimental data. Moreover, in the context of the UQ-VAE with $\lambda < 1$, the true initial pressure images may not available. Extending the UQ-VAE to an experimental setting could be achieved by utilizing transfer learning. In this case, a simulation based dataset is first used to train the network and then subsequently fine tune the network using an experimental dataset consisting of the measurement data, estimate of the measurement noise, and reconstructed photoacoustic images.

	\section{Conclusions}
	\label{sec:Conclusions}
	
	In this work, a neural network based approach for the Bayesian inverse problem of PAT was proposed. The approach is based on the variational autoencoder (VAE) \cite{Kingma2014} and the uncertainty quantification variational autoencoder (UQ-VAE) \cite{Goh2021}. The proposed method was evaluated with numerical simulations using various levels of measurement noise, scaling parameters $\lambda$, and different sensor geometries. The simulations show that the VAE and UQ-VAE frameworks enable rapid and data-consistent reconstruction and uncertainty quantification in PAT. By varying the scaling parameter, the reconstruction performance of the approach can be adjusted, for example in limited view scenarios. Furthermore, the scaling parameter affects on the size and structure of the credibility intervals. It was also shown that the approach is able to respond to variations in data noise levels, and that it can generalize targets outside the training data.

	\bibliographystyle{siamplain}

\end{document}